\documentclass[english,times]{ettauth}
\usepackage{babel}
\usepackage[utf8x]{inputenc}
\usepackage[OT1]{fontenc}
\usepackage{graphicx}
\usepackage{amsmath}

\def\startpage{\pageref{FirstPage}}
\def\endpage{\pageref{LastPage}}
\def\volumeyear{2011}
\def\volumenumber{22}
\def\DOI{ett.1485}

\gdef\copyrightline{Copyright \copyright\ \volumeyear\ John Wiley \& Sons, Ltd.}

\begin{document}

\runningheads{M. Rodr\'{\i}guez P\'erez ET AL.}{Problems in Delay-based
  Congestion Control: a Gallery of Solutions}

\title{Common Problems in Delay-Based Congestion Control Algorithms:\\A
  Gallery of Solutions}

\author{M.~Rodríguez-Pérez\corrauth, S.~Herrería-Alonso, 
  M.~Fernández-Veiga, C.~López-García}

\address{Dept.~Telematics Engineering, E.I.~Telecomunicación,
  Campus~Universitario~Lagoas-Marcosende~s/n, 36310~Vigo,~Spain}

\corraddr{Email address: Miguel.Rodriguez@det.uvigo.es.}

\begin{abstract}
  Although delay-based congestion control protocols such as FAST promise to
  deliver better performance than traditional TCP Reno, they have not yet been
  widely incorporated to the Internet. Several factors have contributed to
  their lack of deployment. Probably, the main contributing factor is that
  they are not able to compete fairly against loss-based congestion control
  protocols. In fact, the transmission rate in equilibrium of delay-based
  approaches is always less than their fair share when they share the network
  with traditional TCP-Reno derivatives, that employ packet losses as their
  congestion signal. There are also other performance impairments caused by
  the sensitivity to errors in the measurement of the congestion signal
  (queuing delay) that reduce the efficiency and the intra-protocol fairness
  of the algorithms. In this paper we report, analyze and discuss some recent
  proposals in the literature to improve the dynamic behavior of delay-based
  congestion control algorithms, and FAST in particular. Coexistence of
  sources reacting differently to congestion, identifying congestion
  appearance in the reverse path and the \emph{persistent congestion} problem
  are the issues specifically addressed.
\end{abstract}

\maketitle

\section{Introduction}
\label{sec:introduction}

Since its early first introduction~\cite{Jacobson96,rfc2581}, congestion
control has been essential in guaranteeing the stability of the Internet and
in enabling its unprecedented growth rate. The original implementation of the
control algorithm used packet loss as the triggering signal that sources are
aware~of to dynamically update the size of their transmission window. More
precisely, the absence of an acknowledgement (ACK) or the arrival of several
duplicate ACKs are regarded as genuine symptoms of congestion, and the source
reacts by shrinking its current window size by half, at least~\cite{rfc5681}.
Over the years, the algorithm has undergone several refinements and
modifications in order to control finely the dynamics of end-to-end flows, but
the fundamental principle of detecting congestion through packet losses
remains untouched in TCP Reno, the most widely deployed TCP version. It is
common knowledge, however, that using packet drops to drive the congestion
control actions exhibits some limitations in wireless
networks~\cite{lassila08:_perfor_tcp_low_wirel_links_delay_spikes} ---where
discarding packets due to transmission errors is not a rare event--- and in
networks with a large bandwidth $\times$ delay product where sharply halving
the window size may be an overreaction. This is the motivation behind a
plethora of protocol proposals that either propose the use of more feedback
information about congestion, like XCP
variants~\cite{Katabi02,xia05:_one_more_bit_enoug,wu09:_effic_fair_explic_conges_contr,almeida10:_explic_conges_contr_based_probab_markin},
or directly advocate the use of congestion signals of a different kind. The
class of delay-based congestion avoidance algorithms (DCA) includes, for
instance, CARD~\cite{Jain89}, DUAL~\cite{Wang92}, Vegas~\cite{Brakmo94},
FAST~\cite{Wei06} and LEDBAT~\cite{Shalunov10}, in chronological order. The
premise of DCA schemes is that \emph{queuing delay variation} is a more robust
way to detect incipient congestion. Consequently, the senders measure
end-to-end queuing delay of data packets and adapt accordingly the
transmission rate, an idea that has also been tried for traffic engineering,
as pointed out by~\cite{alparslan11:_tcp_flow_aware_adapt_path}. There are
also recent examples of mixed DCA and loss-based protocols, such as TCP
Illinois~\cite{Liu06} and compound TCP (CTCP)~\cite{Tan06}, i.e., solutions
using a multiplicity of congestion signals. The latter has been commercially
deployed in the Windows 7 and Vista operating systems TCP stacks, although it
is turned off by default.

Generally, DCAs present several appealing characteristics to support them as
an alternative to the traditional loss-based congestion control protocols.
First, in a homogeneous network environment, one where all sources adapt their
rates to the same congestion signal (queuing delay), the bandwidth in the
bottleneck links is fully utilized, even if \emph{different} source algorithms
are used. The reason is that all DCAs carefully avoid packet losses, so the
instantaneous transmission rate shows less oscillations than in conventional
TCP and TCP-friendly~\cite{Floyd00} sources. Second, the bandwidth is fairly
shared among the competing flows. In fact, DCA sources only react to the
average queuing delay, which is common to every flow in a given
route,\footnote{Clearly, we are not considering the possibility of
  differentiated services in the network.} but not on the propagation delay or
the buffer sizes. In contrast, recall that, in equilibrium, TCP senders with
shorter round-trip times (RTT) attain greater
throughput~\cite{Padhye98a,Shakkotai01}. Third, some DCAs, notably FAST, show
fast convergence times and very small variations in the instantaneous rate,
thus arising as ideally suited for long-fat pipes. Moreover, with the tools of
the network utility maximization
theory~\cite{Wei06,Low01,Kunniyur03,Tang05,Wang05,Choi05,Choi06}, the
equilibrium, fairness and convergence properties of DCAs are now well
characterized and can be carefully engineered.

Despite these clear advantages, researches have also discovered that DCAs can
give rise to a number of anomalous behaviors leading to inefficient or unfair
use of the network resources which, ultimately, have hindered their practical
adoption. Probably, the most dangerous problem is the coexistence of DCAs with
other congestion control protocols. In a heterogeneous network, one where the
flows are responsive to different congestion indications, there may exist
multiple equilibrium points, depending on the network parameters (e.g., buffer
sizes) and the ordering of the flow arrival
times~\cite{Tang05,Tan05,Tang06,Tang07}. This means that the bandwidth share
is, in general, unpredictable and cannot be controllable. Thus, throughput and
fairness may seem chaotic in heterogeneous networks. In many cases, when
confronted to TCP Reno, DCAs receive far less bandwidth than expected, as
loss-based protocols seize the greater part of the bottleneck queue,
see~\cite{Kunniyur03,Low02,Bonal99} for an account. Later, in this paper, we
will describe proposed solutions for controlling inter-protocol fairness
between DCA flows and loss based flows. Note that LEDBAT escapes from this
typical behavior, its main design purpose being able to saturate the
bottleneck, while yielding to standard TCP. A unique network equilibrium
point, optimum in the sense of maximizing a weighted aggregate utility
function, can be enforced by imposing on all the flows two conditions, the
access to a common price and the adoption of a common slow-timescale
adaptation rule~\cite{Tang10}. The use of a weighted utility function implies
that there is some efficiency loss and inter-protocol unfairness, though.

Another impairment is the \emph{persistent congestion
  problem}~\cite{Wei06,Hengartner00,Rperez10}, which is a side effect of the
procedure employed for detecting congestion. For proper operation, DCAs need
to measure queuing delay, and they do this indirectly, usually estimating the
round-trip time and the propagation delay (some newer protocols, like LEDBAT,
estimate directly the one-way delay (OWD)). The queuing delay is the
difference between both, and the propagation delay estimation is simply the
minimum of the observed RTTs. Note that these measurements are a purely local
procedure. As DCA flows maintain a constant amount of traffic (the precise
value is a protocol configuration parameter) queued in the network, newly
established flows are very likely to overestimate the propagation delay and
assume a false available bandwidth. The consequence is a severe unfairness
among the flows. More importantly, this situation tends to persist in time as
long as the mean number of active flows does not vary much.

Finally, a third practical limitation of most DCAs comes from their inability
to distinguish between congestion in the forward path from congestion in the
reverse path. More precisely, when there appears congestion in the return path
from receiver to sender, DCAs wrongly assume that the queuing delay has
increased, and react diminishing their sending
rate~\cite{Fu03,Chan03,Liu05,Herreria07}.\footnote{Obviously, protocols that
  just employ OWD, like LEDBAT, are immune to this problem.} The reduction in
efficiency can be large even if the delay of the ACKs grows only moderately,
but is especially harmful in paths with large bandwidth $\times$ delay
product.

In this paper we will separately analyze each of the three phenomena and
comment several solutions proposed in the literature in the last few
years. The material is not new, but it has appeared scattered in a number of
papers up to now and may not be well known.  We hope that the review can help
to clarify the role of DCAs as well as their potential benefits, and ease the
way for their deployment as general purpose congestion control
mechanisms. Rather than a replacement of current practice in the field of
congestion control, congestion based on delay must be considered as a solution
to coexist for a long time with the binary-feedback loss-based approach.

The rest of this paper is organized as follows. Section~\ref{sec:background}
presents a comprehensive description of FAST as a representative of the
current paradigm of DCA algorithms. In section~\ref{sec:reverse-path-cong} we
present an analysis of the reverse-path congestion problem and several
solutions to it.  Section~\ref{sec:pers-cong-probl} deals with the causes of
the persistent congestion problem and explains how to solve it. The
inter-protocol fairness problem is discussed in
Section~\ref{sec:parameter-tuning} and, like with the two previous problems, a
solution is described. Finally, the conclusions are summarized in
Section~\ref{sec:conclusions}.

\section{FAST: throughput, fairness, stability and optimality}
\label{sec:background}

DCAs turn out to be better than TCP Reno and its variants for data
transmission over large bandwidth$\,\times\,$delay paths, where packet losses
are too scarce to allow the timely adjustment of the sending rate. They also
offer better performance to those applications impaired by sudden changes in
the transmission rate. Their basic assumption is that it is possible to gain
insight into network status observing the variations in the RTT, because the
difference between the RTT and the propagation delay is directly related to
the amount of data enqueued. So, larger differences imply nearness to
congestion. By adjusting the window size based on these variations,
DCA-controlled flows are able to keep a high transmission rate without
inducing congestion. In comparison, TCP Reno (and other protocols with similar
reactions) operate slowly pushing the network to congestion in order to find
out the largest possible window size, namely until a packet loss occurs. The
current window size is then reduced by a multiplicative factor and the
increment phase starts over again.

Since both Vegas and FAST have a similar \emph{modus operandi} (the latter
can actually be modelled as a generalized version of the former~\cite{Wei06}),
and all DCAs display strong similarities among them, we shall describe in this
Section a basic model for FAST, taken as a representative of the entire class
of algorithms. Most of the remarks in the following can also be applied to
other cases with no or only minor adjustments related to the specific utility
function of the protocol.

Like TCP Reno, FAST regulates the window size $w(t)$ in order to adjust the
transmission rate.  At the flow level, the window size varies dynamically as
dictated by the equation
\begin{equation}
  \label{eq:dynamic_fast}
  \dot w(t) = \gamma \alpha \left(
    1 - \frac{q(t) x(t)}{\alpha}
    \right),\,\gamma \in (0,1] \text{ and } \alpha > 0
\end{equation}
where $\gamma$ and $\alpha$ are configuration parameters, $q(t)$ is the
instantaneous queuing delay and $x(t) = \frac{w(t)}{d + q(t)}$ is the
transmission rate, where $d$ denotes the flow's round trip propagation delay.
This equation is just a form of proportional control law, in that the rate of
change in the window size is (multiplied by a scale factor) equal to the
distance to equilibrium $x^\star q^\star = \alpha$. Hereafter, starred symbols
refer to quantities in equilibrium. Proportional control usually yields very
fast convergence times to a stable point, when it exists.

The above dynamic flow level behavior is implemented at the packet level with
the following rule. Every update interval, defined to be a constant time or
some number of RTTs, depending on the specific FAST version, the window size
is updated as
\begin{equation}
  w_{i+1} = \gamma \left( \frac{\hat d w_i}{\hat r} + \alpha \right) +
  (1-\gamma)w_i,\,i=0,1,2,\ldots
  \label{eq:wupdate}
\end{equation}
where $\hat d$ is the current estimation of the round-trip propagation delay,
$\hat r = \hat d + \hat q$ is an estimation of the RTT, and $w_0$ is the
selected initial window size. The accurate estimation of $d$ is a bit tricky,
as it can only be correctly measured in the absence of interfering or
background traffic. In practice, $\hat d$ is set to the minimum RTT observed
during the whole transmission. In the end, this is only a problem when
different FAST flows have different overestimations. As long as all FAST flows
make the same error, the fairness properties are not affected. Later in this
paper, we will study this problem and explain some solutions for it.
Eq.~\eqref{eq:wupdate} brings light to the meaning of the constant $\gamma$.
Actually, it works as a \emph{smoothing factor} or \emph{gain} that controls
the speed of convergence. Its value its taken freely from the interval
$(0,1]$, although $\gamma = 0.5$ is a common choice.

The equilibrium properties of FAST are well established in the
literature~\cite{Wei06,Jin03} and coincide with those of
Vegas~\cite{Low02,Samios03}. Fixing $\dot w(t) = 0$
in~\eqref{eq:dynamic_fast}, it is immediate to see that, under equilibrium,
each flow achieves a throughput
\begin{equation}
  \label{eq:fast-tput-equilibrium}
  x^\star = \frac{\alpha}{q^\star} = \frac{\alpha}{\hat r^\star - \hat d}.
\end{equation}
Note first that, if a set of FAST flows are equally configured (they have the
same $\alpha$) and share a bottleneck link, they all have the same
equilibrium rate $x^\star_i = x^\star$ because the queuing delay $r^\star -
\hat d$ is the same, irrespective of their propagation delays. Secondly, the
physical meaning of $\alpha$ is easily revealed: the product of the queuing
delay by the aggregate transmission rates equals the total amount of data in
transit. Assuming that there is a unique bottleneck link along the path with
bandwidth $C$
\begin{equation}
  \label{eq:enqueued-packets}
  (\hat r^* - \hat d)\sum_{i=1}^n x_i^* = (\hat r^* - \hat d)C = n\alpha,
\end{equation}
for $n$ flows, it follows from~\eqref{eq:enqueued-packets} that each flow
contributes $\alpha$ packets to the bottleneck queue backlog.

There is a trade-off for selecting the proper value of $\alpha$. On the one
hand, we would prefer a small value, to minimize overall latency and buffering
needs in the network. However, this makes it more difficult to accurately
measure the queuing delays, since those will be small too. On the other hand,
large values for $\alpha$ provide faster convergence times.

Considered as a distributed algorithm, the marginal utility function of a FAST
source is the queuing delay $\alpha / x^\star$, whereas it is $\kappa /
(w^\star)^\beta$ for the family of loss-based congestion control protocols,
with $\kappa > 0$ an implementation-dependent constant. For instance, $\beta =
2$ for the classical TCP Reno, $\beta = 1$ for SCTP~\cite{Kelly03} and $\beta
= 1.2$ for HighSpeed TCP~\cite{Floyd03}. The equilibrium point in a
homogeneous network of FAST sources with arbitrary topology is unique and
optimal, i.e., it optimizes aggregate utility. The fairness properties are
immediate from~\eqref{eq:fast-tput-equilibrium} and the global stability of
the algorithm has been proved in~\cite{Choi06} for the case of homogeneous
flows. As expected, not every $(\alpha, \gamma)$ combination produces stable
configurations and several other papers deal with the mathematical conditions
to reach a stable equilibrium, see~\cite{Wang05,Choi05,Tan07}. For conditions
about stability and optimality in heterogeneous networks we refer the reader
to~\cite{Tang10}.

\section{The reverse-path congestion problem}
\label{sec:reverse-path-cong}

Like TCP Reno, FAST assumes that all the congestion it is measuring happens in
the data forward path. However, this assumption is not always true. Congestion
could very well occur in the return path, rendering any reaction to this
congestion futile. In the last years, the number of asymmetric links installed
in the Internet has grown substantially, mainly from residential access lines
(xDSL), increasing the likelihood of congestion in the return path. We will
show that the onset of reverse path congestion can severely degrade the
performance of DCAs. Indeed, the appearance of reverse path congestion has two
direct consequences.

Firstly, it causes some acknowledgement packets to get dropped. TCP sources
will treat these losses like ordinary data packet losses, namely slowing down
their sending rates. The problem affects all TCP variants and is not just a
drawback in FAST. However, it is not particularly harmful unless many
consecutive ACKs are dropped, since TCP ACKs are cumulative and the loss of
one ACK is repaired by the reception of a subsequent one~\cite{Fu03}. The
standardized TCP SACK~\cite{rfc2018} allows selective acknowledgements just
for recovering from multiple lost segments within a RTT. The second
consequence only affects protocols that directly use the variations of the
round-trip time to estimate congestion, like FAST or Vegas. Recall from
Section~\ref{sec:background} that the equilibrium throughput is inversely
proportional to the total queuing delay, cf.~\eqref{eq:fast-tput-equilibrium}.
If we expand $r^\star = q_{\mathrm{f}}^\star + q_{\mathrm{b}}^\star + \hat d$,
where $q^\star_{\mathrm{f}}$ is the (equilibrium) forward queuing delay and
$q^\star_{\mathrm{b}}$ is the (equilibrium) reverse path delay, we get
\begin{equation}
  \label{eq:reverse-explicao}
  x ^\star = \frac{\alpha}{q_{\mathrm{f}}^\star +q_{\mathrm{b}}^\star}.
\end{equation}
From~\eqref{eq:reverse-explicao} it is clear that backward queuing delay has
as much weight as the forward queuing delay to establish the final operating
point, when really only the forward queuing delay should have been taken into
account. At last, this is the one that can be reduced after diminishing the
transmission rate.

It is possible to quantify the effect of backward queuing delay in the
throughput when the propagation delay is $d$. Let us make $d = k
q^\star_\mathrm{f}$ and $\rho = q^\star_{\mathrm{b}} / r^\star$, where $k$ is
a suitable factor. We can solve for $q^\star_{\mathrm{b}}$ to obtain
\begin{equation}
  \label{eq:q_b}
  q^\star_{\mathrm{b}} = (k+1) \frac{\rho\,q^\star_{\mathrm{f}}}{1 - \rho}.
\end{equation}
Substituting~\eqref{eq:q_b} into~\eqref{eq:reverse-explicao} we find that
\begin{equation}
  \label{eq:effect-reverse-final}
  x^* = \frac{\alpha}{q^\star_{\mathrm{f}}} \frac{1 - \rho}{1 + k\rho}.
\end{equation}

\begin{figure}
  \centering
  \includegraphics[width=\columnwidth]{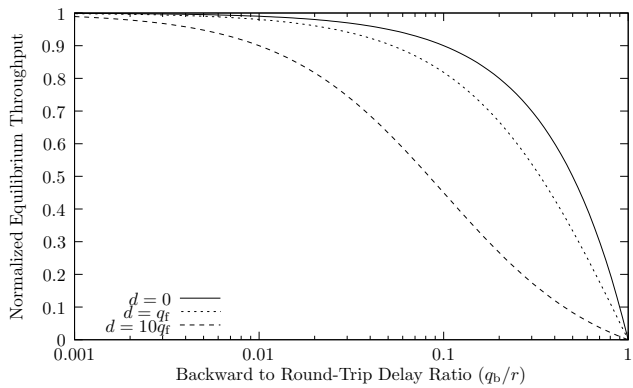}
  \caption{Effect of the backward delay queuing $q_\mathrm b$ on FAST
    throughput for different values of the propagation delay $d$ and a fixed
    forward queuing delay $q_\mathrm f^* > 0$.}
  \label{fig:asymetric-effect}
\end{figure}

In Fig.~\ref{fig:asymetric-effect} we have represented the decay in the
equilibrium throughput of a FAST connection as the backward to total queuing
delay increases, for different round-trip propagation delays.
Obviously, the best case is when the propagation delay is negligible compared
to the queuing delay. Also, the impact on throughput is large for backward
queuing delays greater than around $10\,\%$ of the total delay.

\subsection{Network-assisted solution}
\label{sec:assist-from-netw}

Solutions to the significant fall in performance due to the contribution of
reverse path delay came in two ways, network-assisted and pure end-to-end
approaches.

The most useful technique of the first type is probably~\cite{Liu05}. At
least, it requires the smallest number of changes to the existing network
architecture. Though designed with Vegas in mind, it may be used without
modification in FAST. The authors argue that $r^\star - d$ is not a correct
way to measure queuing delay, since that difference counts the return path
queuing delay too. But, if the delay in the reverse path $r^\ast_\mathrm{b}$
could be measured separately, subtracting it from $r^\star$, a more accurate
estimation $r^\prime = r^\star - r^\star_{\mathrm b}$ allows the source to
adapt its rate correctly.

The idea used in~\cite{Liu05} to calculate $r^\star_{\mathrm b}$ is to install
modified RED~\cite{Floyd93} queues in the return path that mark the packets
with probability
\begin{equation}
  \label{eq:red-reverse}
  p_{\mathrm{ECN}} = \frac{\bigl( b/C - \min_{\mathrm{th}}
    \bigr)^+}{\max_{\mathrm{th}}-\min_{\mathrm{th}}},
\end{equation}
where $b$ is the average queue length, $C$ the link bandwidth and
$\max_{\mathrm{th}}$, $\min_{\mathrm{th}}$ the usual RED thresholds. Now, the
change in $p_\mathrm{ECN}$ is found to be \emph{proportional} to the change of
$r^\star_\mathrm{b}$, so tracking $p_{\mathrm{ECN}}$ is a way of inferring how
much delay contributes the reverse path.

Despite its simplicity, the method has some shortcomings. One is that it
depends on the deployment of modified RED queues on the return path. The
second, and more serious, is that it fails when RED queues are also used in
the forward path. Any other router that also did mark packets would alter the
amount of change in $r^\star_{\mathrm b}$ calculation, effectively making its
value useless. A last drawback is that the new estimation $r^\prime$ does not
eliminate the return path congestion problem in every possible scenario.

\subsection{End-to-end solutions}
\label{sec:end-end-approaches}

Enhanced Vegas~\cite{Chan03} and LEDBAT~\cite{Shalunov10} present alternative
proposals to correct the overestimation of the round-trip time due to the
return path congestion. Unlike other approaches, they are pure end-to-end
solutions requiring no explicit support in the network. Both Enhanced Vegas
and LEDBAT exploit the TCP timestamp options to accomplish their goal. Their
use for LEDBAT is trivial, as it only requires to measure changes in one-way
delay to infer congestion. So, simply adding a timestamp to the data segments
and a measurement result field in the ack packets is sufficient for it.

However both FAST and Vegas employ the queuing delay, and not just its
variations, so the LEDBAT method can not be used. Enhanced Vegas uses the TCP
timestamp options so as to compute an accurate estimation of the backward
queuing delay of the ACK segments. Specifically, the measurement algorithm
estimates the backward trip time $r_\mathrm{b}$ and then deduces the backward
queuing delay from that value. Synchronizing the system clocks of the TCP
sender and receiver is not a requirement for sampling $r_\mathrm{b}$, but
unfortunately aligning both clock speeds is necessary. Otherwise, if the
clocks drift, the measured values become erroneous. The problem may be solved
by resorting to an external (i.e., not part of the transport entities)
procedure to quantify the clock skew itself and compensate for its
value~\cite{Moon99,Zhang02}. Provided this is the case, the backward trip time
is then subtracted from the total RTT $r^\prime = r - r_{\mathrm b}$ exactly
as in~\cite{Liu05}, so that it only includes the term for the forward queuing
delay and the round-trip propagation delay.


While suppressing the backward delay removes the effect of congested reverse
paths, or similarly that of delayed ACKs or asymmetric forward and backward
routes, in most cases, it is not a complete fix, as reported
in~\cite{Herreria07} whose analysis is briefly reproduced here.

Recall~\eqref{eq:wupdate}, the congestion window update in FAST.  Consider for
simplicity, but without loss of generality, the case $\gamma = 1$. In
equilibrium it must hold that
\begin{equation}
  \label{eq:fast-eq-reverse}
  w^\star = \frac{\hat d w^\star}{\hat r^\prime} + \alpha,
\end{equation}
where $\hat r^\prime = \hat r^\star - \hat q^\star_{\mathrm{b}} = \hat
q^\star_{\mathrm f} + \hat d$. Reordering the terms, we have
\begin{equation}
  \label{eq:fast-eq-reverse-reorder}
  w^\star \left(1 - \frac{\hat d}{\hat r^\prime}\right) = \alpha,
\end{equation}
but $1 - \hat d / \hat r^\prime = q^\star_{\mathrm f}/\hat r'$. Using this
identity and~\eqref{eq:fast-eq-reverse-reorder} we obtain the equilibrium
value of the window size as
\begin{equation}
  \label{eq:fast-reverse-w-equilibrium}
  w^\star = \alpha \frac{\hat r^\prime}{\hat q_{\mathrm f}^\star}.
\end{equation}
Finally, from~\eqref{eq:fast-reverse-w-equilibrium} and the well-known
relation $w^\star = x^\star\, r^\star$,
\begin{equation}
  \label{eq:fast-reverse-x-funcof-qf}
  x^* = \frac{\alpha}{\hat q_{\mathrm f}^\star} \left(1 - \frac{\hat q_{\mathrm
        b}^\star}{\hat r^\star}\right).
\end{equation}
Consequently, the term $1 - \hat q_{\mathrm b}^\star / \hat r^\star$ still
introduces a bias on the throughput as the backward congestion increases. This
situation corresponds exactly with the case $d = 0$ depicted previously in
Fig.~\ref{fig:asymetric-effect}. So, either the network-assisted or the
end-to-end solutions discussed above might help to improve throughput in case
of backward congestion, but are effective only if the accumulated queuing
delay in the return path is a small fraction of the total round-trip time.

A simple fix for removing completely the deviation caused by
$q^\star_\mathrm{b}$ was presented in~\cite{Herreria07}. Let $r^*$ stay as is
and modify $d^\prime$ as $d^\prime = \hat d + \hat q^\star_{\mathrm{b}}$. It
is elementary to plug $d^\prime$ and $r^\star$ in~\eqref{eq:fast-eq-reverse}
and find that now
\begin{equation}
  \label{eq:fast-reverse-final-throughput}
  x^\star = \frac{\alpha}{q^\star_{\mathrm f}}
\end{equation}
which is the exact equilibrium rate. In conclusion, \emph{DCAs can be made
  robust against the delay in the reverse path only if the queuing delay of
  ACKs is regarded as part of the propagation delay}.

\section{The persistent congestion problem}
\label{sec:pers-cong-probl}

DCAs are very sensitive to measurement errors in the round-trip time and the
propagation delay. FAST is no exception to this rule. The problem lies in the
way the queuing delay is computed. While the RTT estimation $\hat r$ can be
easily computed by the sender, for each data segment, the value of the
propagation delay is not directly available and has to be inferred somehow.
Most DCA protocols use the same simple heuristic to get $d$, consisting on
identifying $\hat d$ to the minimum $r$ observed throughout a connection
lifetime. This works correctly as long as the intermediate buffers eventually
empty, at all the routers traversed by the packets, and the network path they
follow does not change. In any other case, the propagation delay is
incorrectly measured, and the difference $\hat r - \hat d$ does not truly
represent the queuing delay.

If the network changes the route between sender and receiver, and particularly
when the new path is longer, DCA-controlled flows overestimate the queuing
delay because they record an outdated estimation of $d$. However, it is fair
to say that rerouting does not represent much of a concern in practice, and
that there exist simple ways that most DCA implementations can follow to
ignore the issue. For instance, $\hat d$ can be periodically replaced with the
minimum observed $r$ in the last monitored period. This, at least, limits the
duration of the effects due to rerouting.

On the other hand, ensuring that network buffers get eventually empty and
there are chances for a DCA flow to find out $d$ is more
difficult.\footnote{This possibility should not be overlooked. Consider, for
  example, a homogeneous network with several long-lasting FAST flows. In such
  a scenario, the buffers will not empty, since the flows do always maintain
  $\alpha$ packets queued in the network.} Every time a connection is unable
to get the real propagation delay, because intermediate buffers have some
backlogged data, the queuing delay is underestimated. As a result, the network
path falls into a state of \emph{persistent congestion}, which is the direct
cause of two different problems.

For instance, taking FAST as an example, overestimating the propagation delay
makes FAST flows try to buffer more data in the network than they are allowed
to do. Let $\hat d = k\,d$, with $k > 1$, where $d$ is the real round-trip
propagation delay and $\hat d$ is the measured one, and consider for
simplicity a single FAST flow offering traffic to a bottleneck link with
capacity $C$. From~\eqref{eq:fast-tput-equilibrium} we derive
\begin{equation}
  \label{eq:equilibrium-persistent}
  \alpha = C \left(r^* - \hat d\right).
\end{equation}
Because $k > 1$, $r^* - \hat d < r^* - d$, and since the queue length is
$l = C (r^* - d)$ it is easy to see that $l > C(r^* - \hat d) = \alpha$.

The second problem caused by persistent congestion is worse, in that it leads
to intra-protocol unfairness. Consider $n$ FAST flows arriving consecutively
to a bottleneck link. It is fairly obvious that the latest flow to arrive at
the bottleneck will overestimate its propagation delay, as it will have to
account for the traffic queued by the previous flows. As shown
in~\cite{Rperez10}, in that case the ratio of throughput between the last and
the first flows is, in equilibrium, at worst, $O(n)$.

Although it is arguably not the common pattern that flows arrive consecutively
with the older flows never leaving, the above equation helps to understand the
magnitude of the relative unfairness caused by persistent congestion when a
handful of flows arrive to a shared bottleneck. We will describe next how the
unfairness can be solved.

\subsection{Solving persistent congestion with active queue management}
\label{sec:active-queue-manag}

Similarly to the case of the return path congestion, solutions to persistent
congestion fall in two classes: network-assisted, usually in the form of
active queue management algorithms (AQM), and pure end-to-end approaches. We
will start analyzing AQM proposals first.

In~\cite{La99} routers with RED are proposed in order to allocate bandwidth
more evenly among the flows, regardless of their starting times. However,
finding the appropriate threshold values for the RED gateways is not easy and
remains an open issue~\cite{Alemu04}. Also, this approach does not tackle the
root problem, the incorrect estimation of the propagation delay by the
late-coming flows, but merely address its consequences. Conceptually, RED
breaks the basic assumption of DCAs that packets will not be dropped, since
congestion will not build up. Contrarily, RED drops packets before congestion.
In the end, one could argue that mixing RED and DCA is useless, because one of
the purposes in RED is to warn loss-based congestion controlled sources early
by randomly discarding some packets. DCAs are oblivious to these indications.

The same argument applies to~\cite{Low02}, where the authors suggest a way to
eliminate persistent congestion using REM at the routers.
REM~\cite{Athuraliya01} is an active queue management scheme that keeps buffer
low while sustaining high link utilization. Certainly, with small queues, the
minimum of all measured RTTs is a good approximation to propagation delay, but
the problem now is that there is not enough information in the queuing delay
to enable the detection of congestion. In fact, the price information with REM
is only carried in packet losses, a signal that pure DCAs dismiss. The
necessary modifications in the window adjustment policy of Vegas so that it
interacts adequately with REM are also presented in~\cite{Low02}. Overall, one
could consider this as a mixed DCA- and loss-based protocol, more in the
spirit of CTCP.

In~\cite{Chan04} a new IP option called AQT (Accumulate Queuing Time) is
defined. It is used to collect the queuing time experienced by FAST packets
along a path.  With this scheme, FAST sources must send some probing packets
with the AQT option active, and the routers must compute the queuing time for
each received probing packet and add it to the actual AQT field. As a result,
each connection is able to obtain a good estimate of the propagation delay,
sorting out the queuing time from the RTT measurement. However, the
disadvantages (and complexities) are clear: cooperation from all the routers,
explicit information exchange and modifications of the IP header.

A similar proposal is~\cite{Tan05}, which solves the persistent congestion
problem by marking the ToS field in the IP header with the highest priority
for the first packet of each flow. With priority queuing at the routers, the
highest priority packets will be dispatched immediately even if the router
buffer is not empty and, therefore, FAST will obtain an accurate estimate of
the propagation delay. Despite being simpler than AQT (the ToS field is
standardized), it forces the routers to use priority scheduling or recognize
the packets from FAST flows.

\subsection{End-to-end solutions to persistent congestion}
\label{sec:end-end-solutions}

The need for end-to-end solutions to the persistent congestion problem is
debatable. On the one hand, users can not trust that adequate AQM schemes are
to be widely installed throughout all the Internet as an aid to the use of
DCAs. On the other hand, bottlenecks on the Internet are not very likely to be
monopolized only by DCA flows, instead of containing a mix of both DCA and TCP
Reno data flows. In this latter case, the probability of persistent congestion
is low, given that the bottleneck queues will eventually empty. It is due to
the inherent dynamics of TCP Reno, and it means that DCA-controlled flows will
be able to read the correct round-trip propagation delay. In any case, a
source can find a bottleneck link occupied only by DCA flows, the likeliness
increasing with the proximity of this router to a DCA traffic sender. There
are end-to-end solutions to the persistent congestion problem in those cases
too.

It is also questionable whether a given DCA flow has any interest in helping
to make persistent congestion vanish. After all, late-coming flows greedily
benefit from it, attaining a higher throughput than older flows. We argue,
however, that it is in the interest of flows to collaborate in the avoidance
of persistent congestion. First, persistent congestion causes larger queues at
the bottleneck links, thus larger delays. Secondly, a flow is favoured by the
persistent congestion state as long as it is the latest to become active. This
is a condition that cannot last forever, so a source is likely to take some
actions to counteract the persistent congestion. This gives other flows an
incentive to behave similarly.

The key idea of the end-to-end solutions to the persistent congestion problem
is giving a chance to \emph{new connections} so that the true round-trip
propagation delay can be sampled, without perturbing the flows already
established.\footnote{Another possibility would be that old flows detect the
  arrival of a new flow and react somehow to let the new one seek the right
  propagation delay. As far as we know, this possibility has not been
  explored, however.} Hence, consider as a starting point an ideal situation
where $n$ flows have measured the propagation delay and look for a method
whereby a new flow is able to pinpoint the propagation delay.

The first proposal of such a method appeared in~\cite{Cui06}. Here, the key
idea is to let the new flow pause its transmission sometime after reaching
equilibrium, so as to let the bottleneck queue drain its backlog. It is
assumed that, if the pause is long enough, when the transmission is resumed
the first packet will find an empty queue and will see directly the path's
propagation delay, hence eliminating the persistent congestion bias. However,
the pause length must have an upper bound. The queue only drains until the
rest of the competing flows discover that there is room for more packets, and
then increase their transmission rates. So the queue has to drain completely
in at most one RTT, reducing the generality of this approach. It has been
found~\cite{Rperez08b} that there is a lower bound to the propagation delay
below which the pause is not effective. For a simplified scenario in which all
existing flows have the same propagation delay $d$, the minimum $d$ needed to
empty the queue in just one RTT is
\begin{equation}
  \label{eq:pause-length-critique}
  d > \frac{n \alpha \sqrt{1 + 4n}}{2C} = O\left(n^{\frac{3}{2}} \right),
\end{equation}
where $C$ denotes the bottleneck link's bandwidth. Given the scaling with the
number of flows, pausing transiently would only work in networks with large
propagation delays and a small number of flows.

To overcome these problems,~\cite{Rperez08b} proposes a different behavior for
the new flows. The goal now is obtaining the \emph{error} in the estimation of
the round-trip propagation delay. This error is just the queuing delay due to
the amount of traffic already queued by all the older flows at the bottleneck.
If those older flows all are configured with the same $\alpha$ parameter,
which they should if fairness is desired, this delay amounts to $\epsilon =
\frac{\alpha n}{C}$. The method in~\cite{Rperez08b} estimates indirectly both
$\hat n$ and $\hat C$ to evaluate this error (note that $\alpha$ is already
known). For this it takes advantage of a direct relation between the change in
transmission rate, the measured change in the round-trip time and $n$ if two
conditions hold~\cite{Rperez10}: i) the variation lasts a time short enough so
that the rest of the flows keep their transmission rates unchanged; ii) the
bottleneck queue does not empty during this time. So, provided the two are
met, the last arriving flow can get knowledge of $\hat n$. Once $\hat n$ is
known, the flow can obtain $\hat C$ using~\eqref{eq:fast-tput-equilibrium} and
the relation between $n$ and the queue backlog under persistent congestion,
that is also known~\cite{Rperez10}. Finally, when both $\hat n$ and $\hat C$
are known, the new round-trip propagation delay $d^\prime$ can be set to
\begin{equation}
  \label{eq:error-d}
  d^\prime = \hat d - \frac{\alpha \hat n}{\hat C}.
\end{equation}

The key issue is how to modify the transmission rate. If it increases, there
is a risk to cause packet drops at the bottleneck, which would make the
estimation meaningless. Fortunately, it is very easy to detect such a case.
But if the transmission rate is lowered, the queue can empty all its backlog
and, as discussed before, the measurement fails too. As this second
possibility is much more difficult to ascertain, the authors suggest to modify
the transmission rate only by a small increment. In most situations, there
should be enough room in the bottleneck buffer to hold some more data for just
an RTT, and the failure to do so is not catastrophic.

\begin{figure}
  \centering
  \includegraphics[width=\columnwidth]{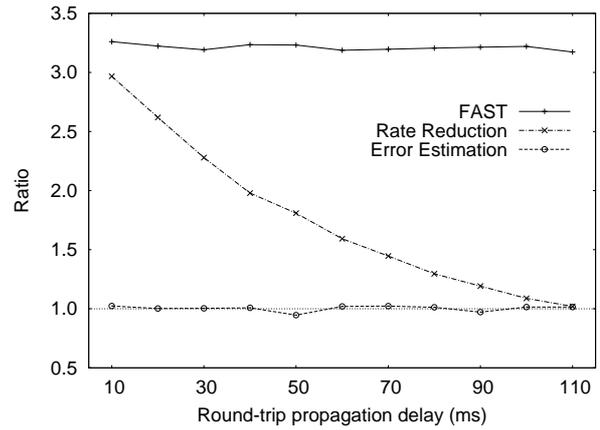}
  \caption{Last arriving flow throughput to previous ones average throughput
    ratio for different persistent congestion avoidance methods with $n=8$.
    Taken from~\cite{Rperez08b}.}
  \label{fig:delay-impact}
\end{figure}
Just to illustrate the effectiveness of the above solutions,
Fig.~\ref{fig:delay-impact} compares the fairness of original FAST against the
solution in~\cite{Cui06}, labeled in the figure as \emph{Rate Reduction}, and
the third approach, labeled \emph{Error Estimation} (EE), for a network with
$n = 8$ flows and different propagation delays. It is clear from the plot that
the EE method works no matter what the real propagation delay is. FAST shows a
strong deviation against old connections, and the rate reduction approach only
works for large propagation delays. Had more flows been set up, fairness among
them would have been achieved only for larger propagation delays~$\sim 
200\,$ms.

\section{The inter-protocol fairness problem or the case of FAST against Reno}
\label{sec:parameter-tuning}

Probably, the best well known problem of DCAs is that they are generally
unable to share the network fairly against the usual TCP variants. This is
also the major problem that prevents the wide-scale deployment of general
purpose DCA algorithms in the Internet, as other advantages, like fewer packet
losses and lower jitter, do not usually pay off receiving less bandwidth. The
description of this inter-protocol unfairness problem is found in several
papers~\cite{Bonal99,Hengartner00,Mo99,Martin03}, although usually only in an
empirical fashion.

A first informal explanation of the root causes of this unfairness appears
in~\cite{Mo99}, referred to Vegas. The translation to FAST is straightforward,
however. Consider the ideal long-run behavior of both FAST and loss-based
protocols, in which the former strive to maintain just $k = \alpha$ packets
(or $\alpha \leq k \leq \beta$ in Vegas), while the latter utilize the full
bottleneck size $B$. The immediate conclusion is that, at any given time, a
FAST flow would have $k$ packets in the bottleneck, while a TCP Reno flow, as
a gross simplification, will have a number in the interval $[0, B - k]$ or, in
average, $(B - k)/2$ packets. So, for two flows sharing a bottleneck, one
running FAST and the other using a Reno-like protocol, their relative
bandwidth share should be
\begin{equation}
  \label{eq:reno-to-FAST-simple}
  \frac{x_{\mathrm{Reno}}^\star}{x_{\mathrm{FAST}}^\star}=\frac{B - k}{2k}.
\end{equation}

\begin{figure}
  \centering
  \includegraphics[width=\columnwidth]{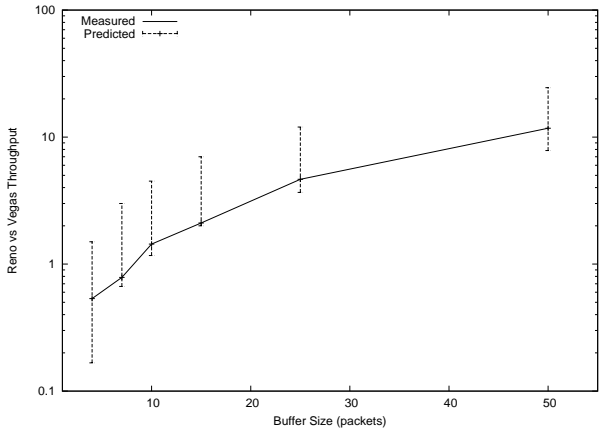}
  \caption{Inter-protocol fairness between TCP-Reno and TCP-Vegas. Original
    data from~\cite{Mo99}.}
  \label{fig:reno-vs-vegas-fairness}
\end{figure}

Notwithstanding its simplicity, the above formula provides acceptable
approximations to the actual performance of TCP Reno when contending with a
DCA. Fig.~\ref{fig:reno-vs-vegas-fairness}
shows simulations results from~\cite{Mo99} where a Vegas flow with $\alpha=1$
and $\beta=3$ shares a bottleneck of variable buffer size with a TCP Reno
flow. Since the exact value of $k$ cannot be predicted with Vegas, the figure
shows both the measured results and the bound~\eqref{eq:reno-to-FAST-simple}
with $k = \alpha$ and $k = \beta$. The results also show that the
inter-protocol fairness of FAST versus TCP Reno depends at least on the buffer
size and the parameter $\alpha$. As expected, larger buffer sizes give more
advantage to TCP Reno flows, which react more aggressively upon a surplus of
bandwidth.

A deeper account of the problem appears in~\cite{Tang05}. In this work the
authors employ a flow model of both TCP Reno and FAST to study the behavior of
a general network where links are potentially shared by both kinds of flows.
The problem is formulated as the maximization of the aggregate utility
functions of the sources. As the marginal utilities of FAST and TCP Reno are
different, and they react to different congestion measures (modeled as link
prices), they can not converge to the same operating point. This is expected,
as it is in full accordance with~\eqref{eq:reno-to-FAST-simple}. Given that
the congestion measures, or link prices, are not the same for FAST and TCP
Reno, in~\cite{Tang05} the authors introduce a \emph{price mapping function}
to feed the same link prices to all sources. This allows to reach two
important conclusions: i) the intra-protocol fairness properties of TCP Reno
and FAST are not affected by the presence of different congestion control
protocols; ii) the inter-protocol fairness properties can be adjusted by
\emph{simply} multiplying the corresponding utility functions by a constant
factor $\mu$ . This clearly opens the door to new Internet congestion control,
as no network changes are needed, not even changes in current congestion
control protocols. Only newly deployed protocols have to be adjusted, by a
constant factor, to be fair against any other in use.

In the case of FAST, whose utility function is
\begin{equation}
  \label{eq:FAST-utility}
  U(x) = \alpha \log(x),
\end{equation}
the modification needed to make it compatible against TCP Reno is trivial.
Multiplying $U(x)$ by a constant factor is directly equivalent to use a new
$\alpha^\prime = \mu \alpha$. The difficulty lies in obtaining this new
$\alpha^\prime$ value in a scalable and decentralized fashion.

A first attempt was presented in~\cite{Tang06}. The long-term average value of
the loss rate is used here to drive the adaption of $\alpha$ to a value such
that FAST competes fairly against TCP Reno. It is then proved that the
equilibrium value of $\alpha$ is given by
\begin{equation}
  \label{eq:alpha-adapted-to-reno}
  \alpha^\star = \frac{q^\star}{\lambda^\star},
\end{equation}
where $q^*$ is the queuing delay at equilibrium and $\lambda^*$ the
corresponding loss rate. It is important to note that the above equation
implies that in a homogeneous network with DCA-controlled flows and with well
dimensioned buffers, $\alpha^\star \rightarrow \infty$ by the lack of packet
losses. The method includes a guard against this risk: $\alpha$ is not
modified at all if there are no packet losses.

Clearly, there will be cases where $\alpha$ is increased from its original
value to attain a fair share of bandwidth against TCP Reno. We could be
tempted to conclude that this means sacrificing the low latency properties of
FAST, as a larger $\alpha$ means more data queued in the network. However,
this is not the case. The total latency remains unaffected, because the larger
$\alpha$ is a consequence of the presence of TCP Reno flows, so the bottleneck
queue was already full before $\alpha$ was increased. Thus, the only effect is
that the buffer share used by each flow at the bottlenecks is modified, but
not its total size. The main penalty of the algorithm is that it only works in
a slow timescale, providing solution to stable scenarios with long-lived FAST
flows and a stable amount of TCP Reno flows. More work is needed to devise new
end-to-end algorithms than can update $\alpha$ in smaller timescales.

Another new approach for improving the coexistence between loss based and DCA
algorithms is presented
in~\cite{budzisz09:_strat_fair_coexis_loss_delay,hayes10:_improv_coexis_loss_toler_delay}.
The respective authors propose methods for the DCA flows to react differently
to queuing delay when there are loss-based flows in the network. For this they
define two operating scenarios: one when the queuing delay is below a certain
threshold and another one when the queuing delay is higher. In the first
scenario, with low queuing delay, they behave like normal DCA flows, trying to
maintain the queuing delay low. However, the more the queuing delay surpasses
the threshold, the less they behave like DCA flows as they begin to
progressively ignore the queuing delay feedback. This lets DCA flows compete
fairly against loss-based flows. The loss-based flows drive the queuing delay
high and DCA flows progressively revert also to loss-based flows. When the
loss-based flows abandon the network, the residual reaction to queuing delay
drive the operating scenario progressively to lower queuing delay points. The
lower the queuing delay, the more the DCA flows respond to queuing delay,
finally putting the operating scenario below the threshold where they behave
like normal DCA flows.

\section{Conclusions}
\label{sec:conclusions}

Although delay based congestion algorithms, like those present in Vegas or
FAST have long promised better performance that TCP Reno, they have not been
deployed in the Internet for several reasons: overreaction to congestion in
the return path, intra-protocol fairness problems in networks with persistent
congestion and inter-protocol fairness among heterogeneous flows.

In this paper we have reviewed state-of-the-art solutions to all these
problems, some requiring changes in the network routers and others consisting
purely in end-to-end approaches. We have presented, at least, one end-to-end
solution to each of the aforementioned problems.

Our overall consideration is that improved versions of DCAs are ready to be
used in the Internet and coexist with the classic loss-based congestion
control algorithms. For instance, the fix to the persistent congestion problem
produces some variations in the sending rate of the aggregate FAST traffic,
but these variations are much smaller than typical rate variations of TCP Reno
traffic. Also, the fix for the inter-protocol fairness increases the amount of
traffic that FAST sources queue at routers, but only when confronted with TCP
Reno and recall that even in that case the end-to-end delay does not get
larger because the total queue length does not change.

\acks{This work was supported by the ``Ministerio de Ciencia e Innovación''
    through the project TEC2009-12135 of the ``Plan Nacional de I+D+I''
    (partially financed with FEDER funds).}

\bibliographystyle{wileyj}
\bibliography{IEEEfull,biblio}
\end{document}